# The Research Production of Nations and Departments: A Statistical Model for the Share of Publications[1]

Mike Thelwall, Ruth Fairclough
Statistical Cybermetrics Research Group, University of Wolverhampton, UK.

Policy makers and managers sometimes assess the share of research produced by a group (country, department, institution). This takes the form of the percentage of publications in a journal, field or broad area that has been published by the group. This quantity is affected by essentially random influences that obscure underlying changes over time and differences between groups. A model of research production is needed to help identify whether differences between two shares indicate underlying differences. This article introduces a simple production model for indicators that report the share of the world's output in a journal or subject category, assuming that every new article has the same probability to be authored by a given group. With this assumption, confidence limits can be calculated for the underlying production capability (i.e., probability to publish). The results of a time series analysis of national contributions to 36 large monodisciplinary journals 1996-2016 are broadly consistent with this hypothesis. Follow up tests of countries and institutions in 26 Scopus subject categories support the conclusions but highlight the importance of ensuring consistent subject category coverage.

## 1 Introduction

Research assessment is an increasingly important activity, from individual researcher self-evaluations (Wouters & Costas, 2012) to national merit-based exercises to allocate funding (e.g., Butler, 2008). For this, both the quantity and quality of outputs may be considered. For quantity, the number of outputs might be counted whereas peer review or citation-based indicators may inform quality judgements. One of the attractions of the h-index is that it combines both, albeit in a controversial way (e.g., Costas & Bordons, 2007). The main approach seems to be to largely ignore quantity and focus on quality. In some national research evaluation exercises, there is a quantity threshold, such as four journal articles per scholar in six years, but evaluating quality alone (Franceschet & Costantini, 2011; Wilsdon, Allen, Belfiore, Campbell, Curry, et al., 2015). Nevertheless, quantity information is essential to give a rounded impression of the outputs of a set of researchers, from individual research groups to countries. Such statistics are often reported at the national level (Elsevier, 2013) but are rarely analysed in terms of underlying capacity. There is thus a need for a more theoretical analysis of research production, in the sense of the capacity to produce outputs, to help interpret the statistics reported.

### *1.1 Research production indicators and applications*

Research production is usually equated with the number of publications produced of a given type or set of types, although it is sometimes analysed in a way that includes impact, efficiency or quality components (Abramo & D'Angelo, 2014). There have been previous attempts to statistically analyse research production to find its determinants. For individual scientists, it has been shown to relate to incentives (Levin & Stephan, 1991), motivation (Taylor, Locke, Lee, & Gist, 1984), institutional environment (Dundar & Lewis, 1998), collaboration (Abramo, D'Angelo, & Di Costa, 2009), internet use (Barjak, 2006), academic rank (Abramo, D'Angelo, & Di Costa, 2011) and gender or other personal factors (Sax, Hagedorn, Arredondo, & Di Crisi, 2002). These studies have taken the statistical assumption that there are both systematic and uncontrolled factors but have not specified a basic research production model. Research productivity in the sense of efficiency has also been investigated by comparing outputs to inputs (Abramo, D'Angelo, & Pugini, 2008; Abramo, Costa, & D'Angelo, 2015). Production in the sense of the probability of an author to select a given journal has also been tested against Lotka's law (Rowlands, 2005). From a distribution perspective, the number of publications from a set of research groups may follow a lognormal distribution (a visual observation based on Figure 1 of: van Raan, 2006), suggesting a degree of statistical regularity at the aggregate level, so that the existence of an underlying model is plausible. Despite all these studies and one model for individual researchers (Koski, Sandström, & Sandström, 2016) there does not seem to have been a previous attempt to use publication counts to model the underlying production of a group, in the sense of its capacity to produce research outputs.

Governments assessing national research performance often commission reports that analyse, amongst other factors, the national *share* of the world's publications. A UK commissioned report, for example, includes shares of the world's Scopus-indexed articles, citations and highly cited papers in addition to the same values per researcher and per unit research expenditure (Figure 1.1 of Elsevier, 2013), as well as diagrams showing changes over time in the relationship between article share and average impact (Figure 1.2 of Elsevier, 2013). A U.S. National Science Foundation report includes, amongst many other statistics, 50 countries' shares of the world's publications (Table 5-23 of NSF, 2016) and a graph showing changes in national shares of publications (Scopus-indexed books, conference papers and journal articles) over time (Figure 5-24 of NSF, 2016). An OECD report includes, amongst other figures, national shares of publications (Figure 1.2 of OECD & SCImago, 2016). This document also reports publications per million inhabitants. These reports assume that analysing publication share is meaningful as part of a set of indicators. Other national-level reports include total output indicators but no output share indicators, such as one for the European Commission (Science-Metrix, 2015).

Shares of publications have also been analysed at lower levels of aggregation. A report for the U.S. National Science Foundation analysed changes over time in the share of articles produced by different sectors of the economy (Figure 11 of Côté, Roberge, & Archambault, 2016). This report also compared absolute numbers of publications at the national level. Sector shares within the national output are also reported in the New Zealand research performance report (Ministry of Business, Innovation and Employment, 2016). This report primarily compares New Zealand to similar countries rather than the rest of the world.

## *1.2 The research production capability model*

This paper introduces a model of underlying production for a group of researchers that is based on their *share* of the world's output. From the perspective of a small research group, a more natural model would be for the total number of articles published by them. Using share rather than volume is nevertheless appropriate because the number of outputs per researcher may alter over time in line with changes in technology and publishing opportunities. For example, if a journal's coverage doubled (or a Scopus subject category doubled in size) then it would not be reasonable to *expect* a group's output to double in response. Nevertheless, if a journal doubles in size then unless it attracts new contributors for the expanded coverage (e.g., because its scope also expands) then the same contributors must fill its pages by contributing, on average, the same share as before the expansion. Expansions are likely to be triggered by a backlog due to increased demand or an attempt to take an increased share of articles from other similar journals. Thus, it seems reasonable to hypothesise that the probability of publishing in a journal is more fundamental than the number of articles published in that journal, or at least a reasonable alternative perspective.

The same argument applies to entire subject areas to a lesser extent. If a subject area expands by some of its journals getting larger, then the same set of authors would presumably be expected to write the extra articles. Nevertheless, if a citation index increases its coverage of a subject area by adding new journals then, since academic journals can have a specialist audience that differs to some extent from all other journals, they are likely to bring at least a few new specialist authors with them. Thus, the share of the publications of the existing groups within subject categories seems likely to decline (e.g., see: Abramo, D'Angelo, & Di Costa, 2008). This will also occur of the citation index adds journals with a national focus, such as a set from China. The probability argument is therefore much weaker for subject areas unless they are pre-filtered for journal changes, except for periods in which their constituent journals are unchanged.

At the system level, the amount of research published, or at least being indexed, is expanding. In Scopus, for example, the number of journal articles has increased rapidly since 1943 (search for DOCTYPE(ar) and click *Analyse search results*). This is probably partly due to increases in the number active researchers, especially in countries like China, India and Brazil. Thus, most countries seem to almost continually increase their number of outputs and so increases are "normal", complicating the task of detecting significant underlying changes based on volume alone.

The model of production capability makes the simplifying assumption that at the start of the year, each article published in a journal or subject area within that year has the same (unknown) probability $\Pi$ to be published by a specified group (e.g., department or country). With this assumption, the number of articles out of the total $n$ published in the journal at the end of the year follows the binomial distribution $B(n, \Pi)$. If a proportion $p$ of the articles in the subject or journal are published by the specified group, then $p$ is an estimator for $\Pi$ and a 95% confidence interval can be calculated for $\Pi$ using Wilson's (1927) score interval.

$$\frac{p + \frac{1.96^2}{2n} \pm 1.96\sqrt{\frac{p(1-p)}{n} + \frac{1.96^2}{4n^2}}}{1 + 1.96^2/n} \tag{1}$$

The simplifying assumption above is not an intuitive way of thinking about research publishing and does not seem to have been suggested before. It is more intuitive to think in terms of random factors affecting the total number of publications produced by a group. Nevertheless, the above model fits naturally with indicators reporting publication share and is designed to give a simple way to help interpret their values. The result of the process is more intuitive: confidence limits for the *share* of publications produced by a group.

The above probability relates to a consideration of the total number of publications produced by a group together with a system expansion assumption. Suppose that group $g$ produces $n_{g0}$ publications in year 0 out of $N_0$ publications in a subject area or journal. Suppose that a few years later the system has expanded to $N_i$ publications due to greater ease of publishing, an increase in the number and size of journals or expanded coverage of the citation index used. If group g's capacity to produce research has stayed the same, then the number of publications it produces in year $i$ might be expected to rise in line with the world to:

$$n_{gi} = n_{g0}\frac{N_i}{N_0} \qquad (2)$$

Dividing both sides by $N_i$ gives:

$$\frac{n_{gi}}{N_i} = \frac{n_{g0}}{N_0} \qquad (3)$$

Thus, the unchanging factor is the proportion of the world's publications produced, which is the value that the probability $p$ in the model of production capability is estimating. From the number of publications perspective, however, there is no need to assume that there is an underlying probability $p$, but the number produced could be modelled by any suitable discrete distribution. Of these, the production capability model is the simplest and therefor the logical first choice in the absence of evidence to the contrary.

Simplifying models have previously been used to explain empirical citation distributions (e.g., de Solla Price, 1976) and this strategy can support better understanding and more accurate, analyses, despite the simplifications being clearly incorrect on a small scale. For example, the Matthew effect, when turned into a mathematical model, assumes that the proximity of an article being cited by a new publication is approximately proportional to its existing citation count (de Solla Price, 1976; Merton, 1968), despite scientists never choosing papers to cite in this way (although citation counts can influence their choice).

From a statistical perspective, the model of production capability turns populations into samples, making statistical inference possible. Without a theoretical perspective if a group publishes 2% of the articles in a year then this is a precise figure, rather than a sample from a larger population. If the group publishes 2.1% of the journal's articles the following year, then there is no way to decide whether this increase is big enough to reflect an underlying change. With the model of production capability, the 2% can be used to estimate the underlying probability to publish in the journal (0.02 in this case) together with confidence intervals. These can be used to decide whether the subsequent year's value (0.21) is different enough to suggest that the underlying probability to publish has changed. This is an example of the social science apparent populations strategy that regards complete data sets as samples from the set of theoretically possible outcomes from the environment (Berk, Western, & Weiss, 1995).

From the perspective of those constructing research indicators, the production capability model is not intended to replace all other publication output indicators but is for when share-based indicators are employed, as in the examples given above (Elsevier, 2013; NSF, 2016; OECD & SCImago, 2016; Côté, Roberge, & Archambault, 2016). This is normally in conduction with related data, such as absolute numbers of publications.

## *1.3 Prediction intervals*

Confidence intervals, such as (1), are useful to delimit the underlying probability to publish but it may also be useful to have a test to detect whether this probability is likely to have changed, based upon a second data set, such as results from a second year. Standard confidence intervals cannot be used for this since they delimit the underlying population parameter rather than future sample proportions. It is therefore necessary to switch to prediction intervals, which give an interval for future sample proportions from the same population. The following formula is a 95% prediction interval for the number of publications $\hat{Y}$ (out of $m$) being published by a group in a subsequent year, if it published $X$ out of $n$ in a previous year (joint distribution method, equation 12 in: Krishnamoorthy & Peng, 2011).

$$(L,U) = \frac{\hat{Y}\left(1 - \frac{1.96^2}{m+n}\right) + \frac{1.96^2 m}{2n} \pm 1.96\sqrt{\hat{Y}\left(m - \hat{Y}\right)\left(\frac{1}{m} + \frac{1}{n}\right) + \frac{1.96^2 m^2}{4n^2}}}{1 + \frac{1.96^2 m}{n(M+n)}} \quad (4)$$

This prediction interval is for the number of publications produced by the group in the second sample. L can be rounded up, with U rounded down to whole numbers. A prediction interval for the proportion of articles published by the group in the second time period is therefore:

$$(L/m, U/m) \quad (5)$$

If the total number of publications produced by the world set (journal, category or Scopus/WoS database) is approximately constant then it would be reasonable to set $m = n$ so that the prediction intervals could be calculated in advance.

## *1.4 Comparison with alternative strategies*

In comparison to efficiency models of research that consider the ratio of system outputs to system inputs (Abramo, D'Angelo, & Pugini, 2008; Abramo, Costa, & D'Angelo, 2015), the proposed model has the advantage that it requires only (currently) easily available data but the disadvantage that it delivers less useful information to the research manager. Another model focuses on the output per researcher, as reflected by a publication database. It takes into account that the most unproductive researchers are ignored in many bibliometric analyses because they have no database records, giving misleading national average papers-per-researcher indicators (Koski, Sandström, & Sandström, 2016). This models a group as having both publishing and non-publishing researchers and uses a statistical distribution assumption to accommodate the missing data. This productivity (papers per researcher) approach is complementary to the production (share) approach of the current paper.

There are other ways of dealing with uncertainty in bibliometric data. The most widely used at a practical level is probably common sense or expert judgement of the research manager considering multiple indicators with their own background knowledge of the local and world research context. This takes a theory-free approach, which has the drawback that different people may interpret the results in different ways and the decision making is not fully transparent.

A variety of strategies have been proposed support or replace expert judgement. Funnel plots for universities are a graphic device to help managers to compare productivities by visually illustrating uncertainty (Abramo, D'Angelo, Grilli, 2016; Abramo, D'Angelo, Grilli, 2015). This approach has the advantage that it gives an overview of multiple institutions in a single graphic that combines indicator values and uncertainty information. This device uses traditional confidence intervals but could also use the confidence or prediction interval approach introduced here. A different tool, stability intervals, are like confidence intervals but deal with uncertainty in derived from the publication dataset used, employing bootstrapping (Waltman, Calero-Medina, Kosten, et al., 2012). This is a less theoretical alternative to the current approach but its results are probably similar.

Finally, and related to the previous point, the use of statistical inference in scientometrics has been challenged, for example on the basis that there are many causes of randomness and that statistical inference (typically for citation indicators but many arguments also apply to publication indicators) is therefore rarely well founded (Waltman, 2016). The new model attempts to reduce, but not eliminate, the ambiguity in the causes of randomness by proposing a simple model.

# 2 Research questions

It is impossible to directly check the model of production capability and so the research questions are pragmatic in the sense of assessing its results: the extent to which it gives plausible information. Journals are used to test the model to avoid the results being affected by changes in subject area coverage over time. The second research question addresses whether the results are likely to be affected by international collaboration, which complicates the issue of assessing production capability. The third and fourth research questions are more exploratory. They assess the two contexts in which the model may be applied in practice but for which it is difficult to produce long term robust tests.

1. Do prediction intervals for the proportion of a country's articles within a journal give plausible results in the sense of close to 95% of following year proportions falling within them and the proportion decaying for subsequent years?
2. Are prediction intervals for the proportion of a country's articles within a journal more stable if articles are considered with all authors from the country rather than with any author from the country?
3. Do prediction intervals for the proportion of a country's articles within a Scopus subject category give plausible results in the sense of close to 95% of following year proportions falling within them and the proportion decaying for subsequent years?
4. Do prediction intervals for the proportion of an institution's articles within a Scopus subject category give plausible results in the sense of close to 95% of following year proportions falling within them and the proportion decaying for subsequent years?

# 3 Methods

## *3.1 First research question: Journal-level analysis*

This paper recycles a set of 36 large monodisciplinary journals (Appendix A) from a previous study of citation prediction interval accuracy and uses a similar strategy, but on a different aspect of the data (publication counts rather than citation counts) and for different goals (production capability modelling rather than field normalised prediction interval accuracy validation) (Thelwall, in press). This set was formed by taking the 50 current journals in Scopus with the most refereed journal articles 1996 (to ensure long term large data sets) excluding non-scientific journals (e.g., Jane's Defence Weekly) and general journals (e.g., Nature, Science, PLOS ONE) because these publish different subject areas. Journals that did not publish continuously 1996-2016 were also excluded. Large journals are important to give statistical power to the analysis and the long period was needed to assess differences over time. The data was gathered in February and March 2017 and so should be close to complete for all years.

The ten countries with the most articles in the combined set of 36 large monodisciplinary journals was assessed for the plausibility of prediction intervals for their probability of authoring articles in each journal (RQ1). Countries were selected rather than research groups or institutions for increased statistical power and long term stability. Individual research groups sometimes have policy shifts that result in increased or decreased submissions to specific journal (e.g., Howard, Cole, & Maxwell, 1987), which would undermine the findings.

A 95% prediction interval based on the underlying probability to author a paper in the journal was calculated for each country, journal and year except 2016 (n=36 x 10 x 20=7200 prediction intervals) using (5) above. For each year except 1996, the proportion of articles with at least one author from a given country was also calculated (n=36 x 10 x 20=7200 sample proportions). For each journal and country, all sample proportions were checked to see whether they occurred within the prediction intervals for each preceding year, recording this against the time lag between the prediction interval and subsequent year. For example, China published 8 out of 2201 articles in the Journal of the American Chemical Society in 1996, with a 95% prediction interval of (0.0018,0.0072). In 1997 (1-year gap), it published 11 out of 1957, which is 0.0056, within the 95% prediction interval. In contrast, in 2000 (4-year gap), it published 14 out of 1396 articles and the proportion 0.001 falls outside the 1996 prediction interval. This last check was performed for all pairs of first and subsequent years except for the pairing 1996 and 2016 (20-year gap) because this gave a small sample size (n=36 tests in contrast to between n=72 and n=720) for the remaining year gaps).

## *3.2 Second research question: National research*

The above steps were repeated for articles with all authors from each of the ten selected countries, excluding internationally collaborative research. For example, in the second test only articles with all authors from the USA were included. The basic data for this paper is therefore the outcomes of 150,480 tests for whether the proportion of articles in a year and journal falls within a 95% prediction interval for the same country and journal and a previous year. These tests were aggregated by country and year gap to reveal trends.

## *3.3 Third research question: Scopus categories instead of journals*

For the third research question, a set of journal articles from 26 subject categories in Scopus from 1996 to 2015 was recycled from a previous citation analysis paper (Appendix B). These are the 7$^{th}$ categories within the Scopus broad categories. For example, Forestry was selected from the broad category Agricultural and Biological Sciences. Each subject category and year combination has a maximum of 10,000 journal articles, a system limitation. When there were more than 10000 articles, the first and last 5000 were included. This is likely to affect the results for the larger categories because, depending on annual differences in indexing, there may be journals with national specialisms that are less comprehensively included in the subject category in some years. The methods above for journal articles were replicated for this set of subject categories.

## *3.4 Fourth research question: Institutions instead of countries*

For the final research question, the 26 subject categories in Scopus from 1996 to 2015 were reused at the level of institution, choosing universities from the THES ranking (THES, 2017) with ranks 1, 10, 20, 40, 80, 160 and 320 for a range of different institution levels.

Tests were only conducted for pairs of years when there was at least one article published by the group in the first year to avoid a situation with many positive results by default due to no publishing at all in the category. This mostly affected the data for the fourth research question.

Appendix C contains instructions for the above with the free software Webometric Analyst.

# 4 Results and discussion

The first two research questions are grouped together in the results, investigating the frequency with which publication shares for a country in a journal fall within 95% prediction intervals from previous years.

## *4.1 Journals*

After a one year gap, between 75% (China) and 92% (Canada) of the proportions of publications in a journal from a country fall within a 95% prediction interval calculated from the previous year (Figure 1, 2: 1-year gap). If the statistical model is true and the underlying probability of publishing each article in each journal did not change in a 1 year period then the percentage of sample proportions from the second year falling within 95% confidence intervals from the first year would be close to 95%. Thus, figures substantially above 95% in the graph would suggest a violation of the model and less variability than would be predicted by it. Since the figures are below or substantially below 95%, they are consistent with the model, although the size of the gap below 95% in Figures 1 and 2 is probably due to changes in the underlying probability at some or all points in time during 1996-2016 for all countries and journals.

To set the results in context, if the data in Figures 1 and 2 had been substantially above 95% at any point then this would have shown that the underlying assumptions were implausible (but, as discussed below, this does not apply to small sample sizes and low probabilities). Conversely, much lower figures, perhaps under 50% for a gap of 1, would suggest that the prediction intervals were not useful.

As the time gap between the first year (prediction interval) and second year (proportion tested) increases, the percentages fall steadily to below 70% for all countries (Figures 1, 2). The falling proportion as the gap increases is broadly consistent with the statistical model on the basis that country proportions often experience trends and larger gaps therefore associate with a smaller proportion falling within a prediction interval calculated from the start of the gap.

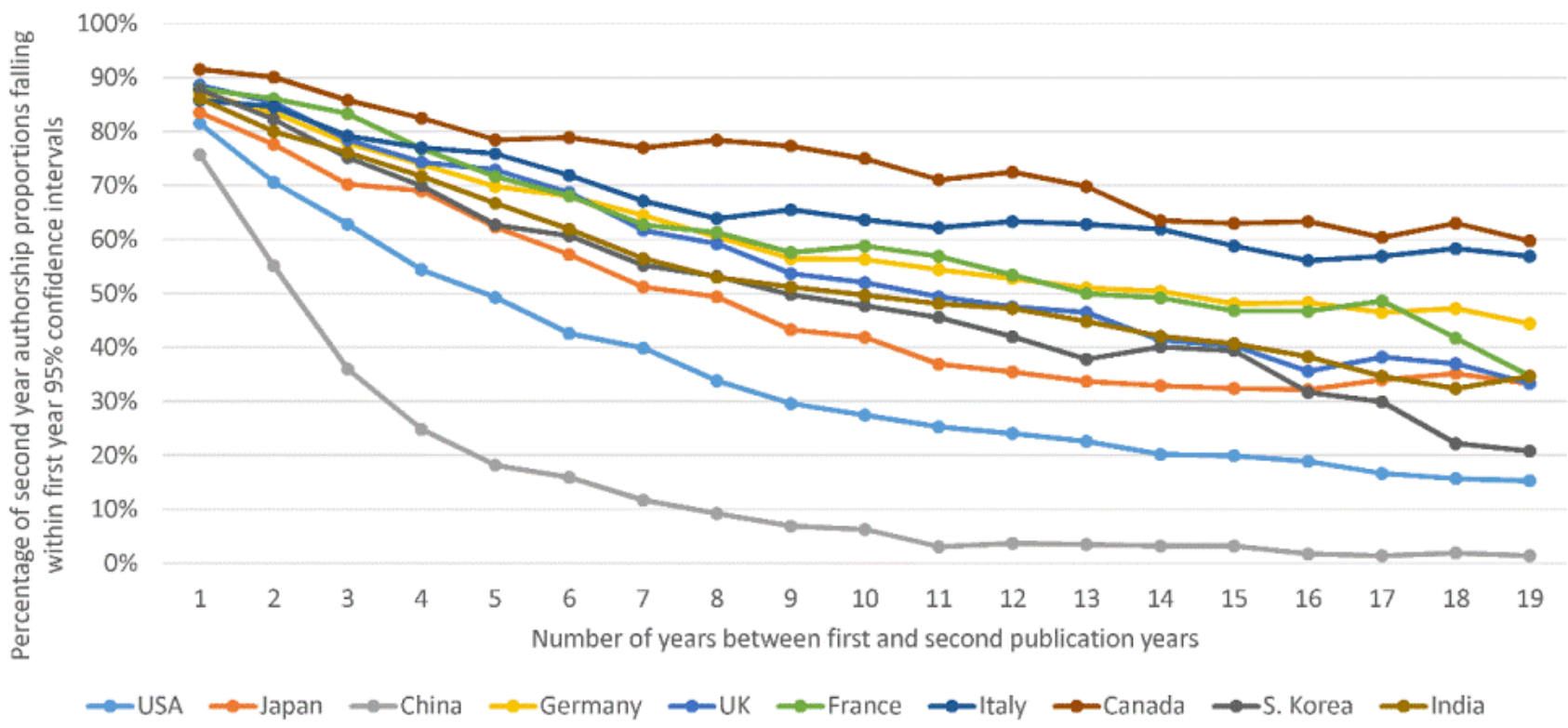

Figure 1. The percentages of times that the proportion of articles published by a country within a journal falls within 95% prediction intervals from previous years, against the gap between the two years. Data for each country is for articles with any author from that country. There are 720 journal/year pairs for a 1 year gap falling linearly to 72 journal/year pairs for a 19-year gap. The raw data comprises n=75,240 individual tests.

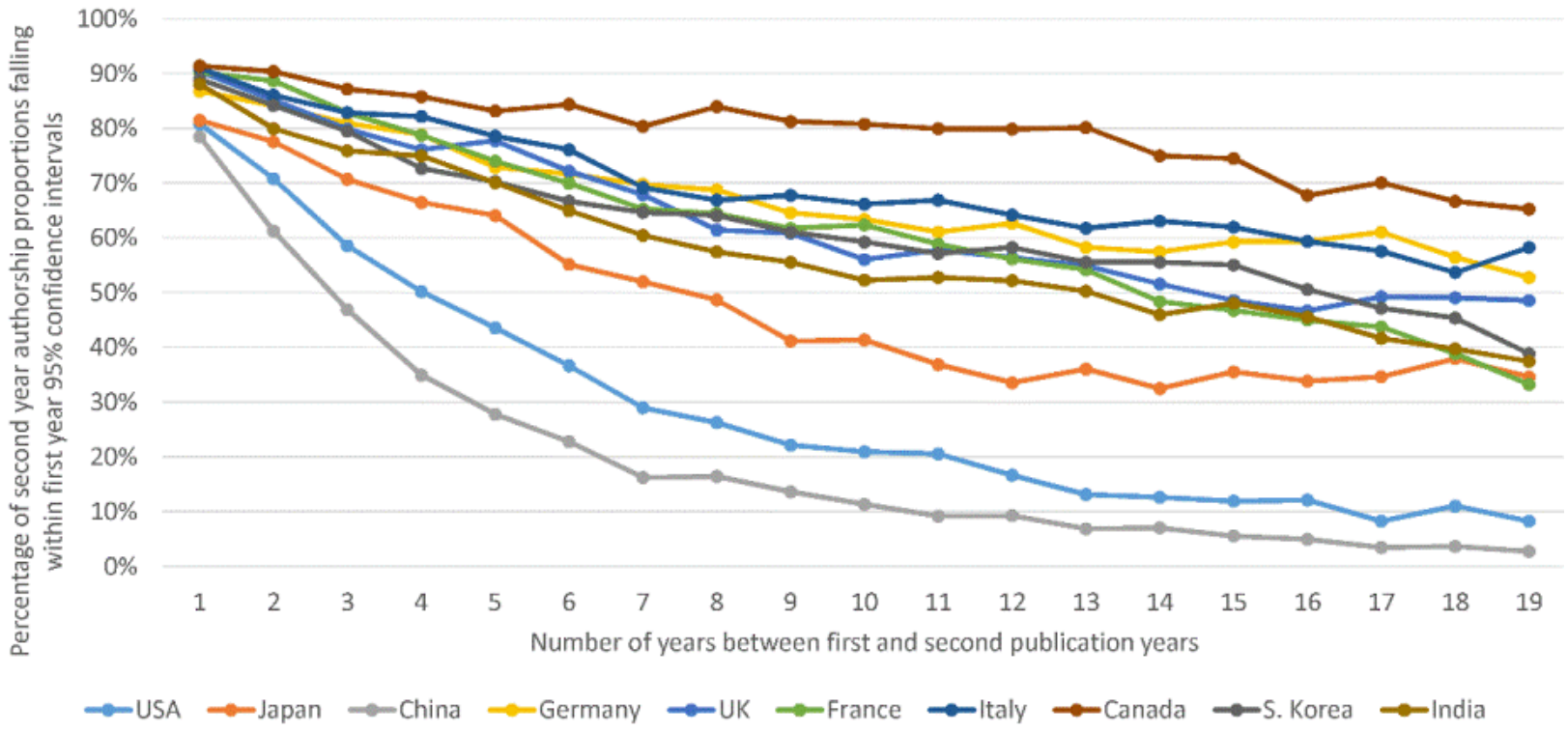

Figure 2. As Figure 1 except that the data for each country is for articles with *all* authors from that country.

In answer to the second research question (Figure 3), for some countries, prediction intervals are more likely to contain subsequent proportions if articles with any author from the country is considered (e.g., USA) and for others the opposite is true (e.g., South Korea). Nevertheless, the difference is not large for the shortest gaps.

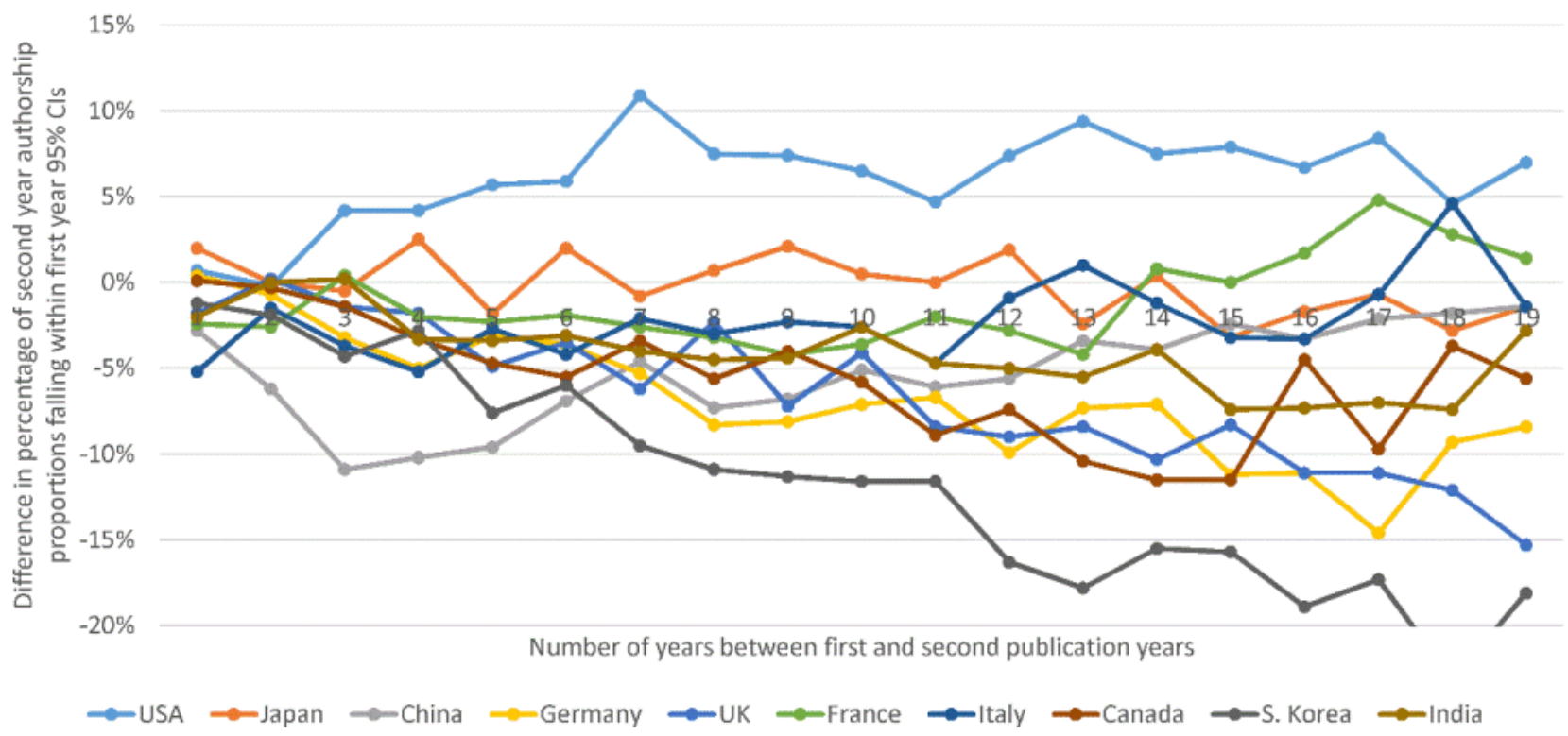

Figure 3. Figure 2 subtract Figure 1. The height of the line is the extent to which prediction intervals for articles with any authors from the stated country are more likely to contain subsequent proportions than prediction intervals for articles with only authors from the country.

Within the broad findings, the trends for individual countries vary. Considering Journal of the American Chemical Society, for example, the proportion of articles with authors from China increased almost every year from 1996 to 2016 (Figure 4). The years with small increases or decreases are consistent with this trend, as can be seen from the prediction intervals: A smooth monotone increasing line could be drawn on the graph through all of them. Thus, policymakers should not be concerned about the deviations from the overall monotone increase because these could be due to natural random variations.

In the opposite trend direction, the USA's share of Journal of the American Chemical Society articles decreased steadily over time 1996-2016 (Figure 5) and even the largest deviation from this, the increase in 1999, is consistent with no underlying growth because it is possible to draw a monotone decreasing line through all the prediction intervals. Thus, whilst something unusual might have happened in 1998 (e.g., a large special issue with a topic of interest mainly outside the USA) or 1999, this is not a necessary conclusion from the data.

A different case is India, with a much lower proportion of Journal of the American Chemical Society articles (Figure 6). The prediction intervals are relatively large for India compared to the line height, as can be seen by comparing Figures 4-7. The larger line fluctuations match the wider prediction intervals.

Canada (Figure 7) is the most stable country overall in the main data (the highest line in Figures 1,2). This is reflected in a reasonably constant share of publications, albeit with a gradual decrease until 2013, followed by a sharper decrease. For Figures 4-7, when a point does not lie within a prior prediction interval then there is evidence of an underlying change in the sense that a hypothesis test at the 95% level would reject the null hypothesis that the two proportions were from populations with the same underlying (i.e., population) probability to author each article in the journal (this is different from how confidence interval overlaps should be interpreted). A combination of this information and the trend gives strong evidence that there has been a long term decrease in the Canadian probability to publish in the Journal of the American Chemical Society.

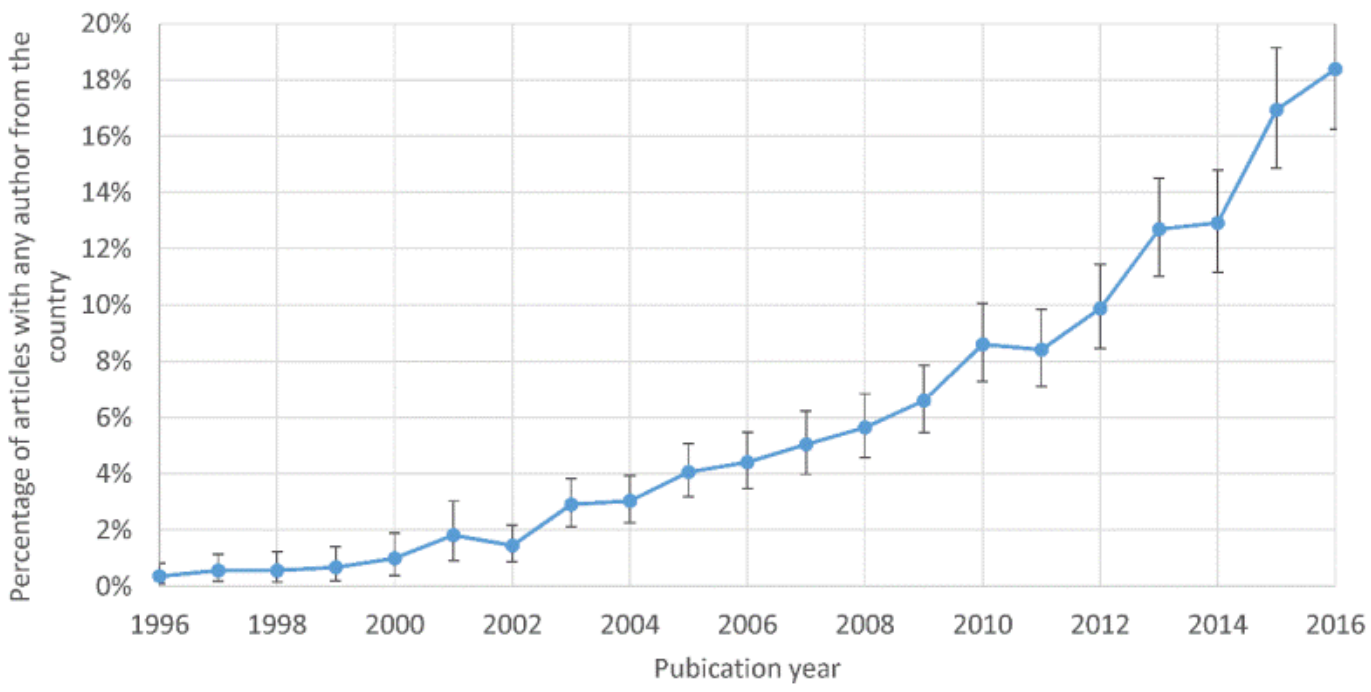

Figure 4. Prediction intervals for the probability (expressed as a percentage) that each article in Journal of the American Chemical Society has at least one author from China. Each prediction interval assumes that the overall number of articles in the journal does not change over time.

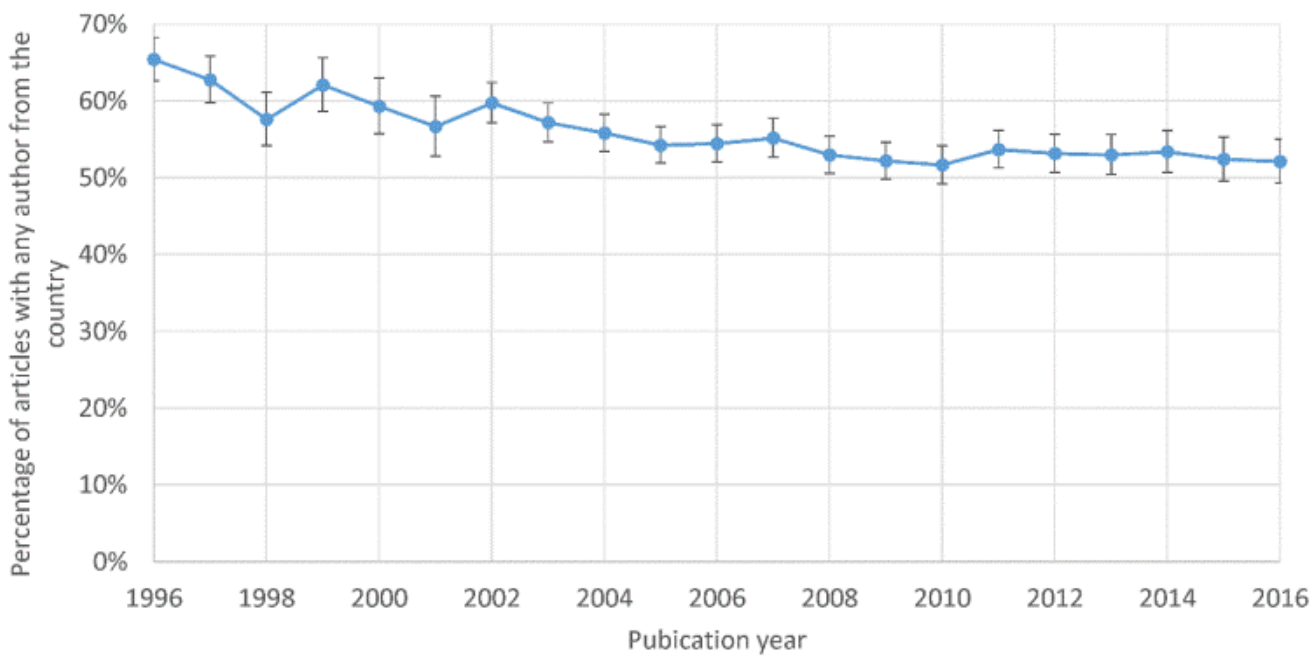

Figure 5. As Figure 4 for the USA.

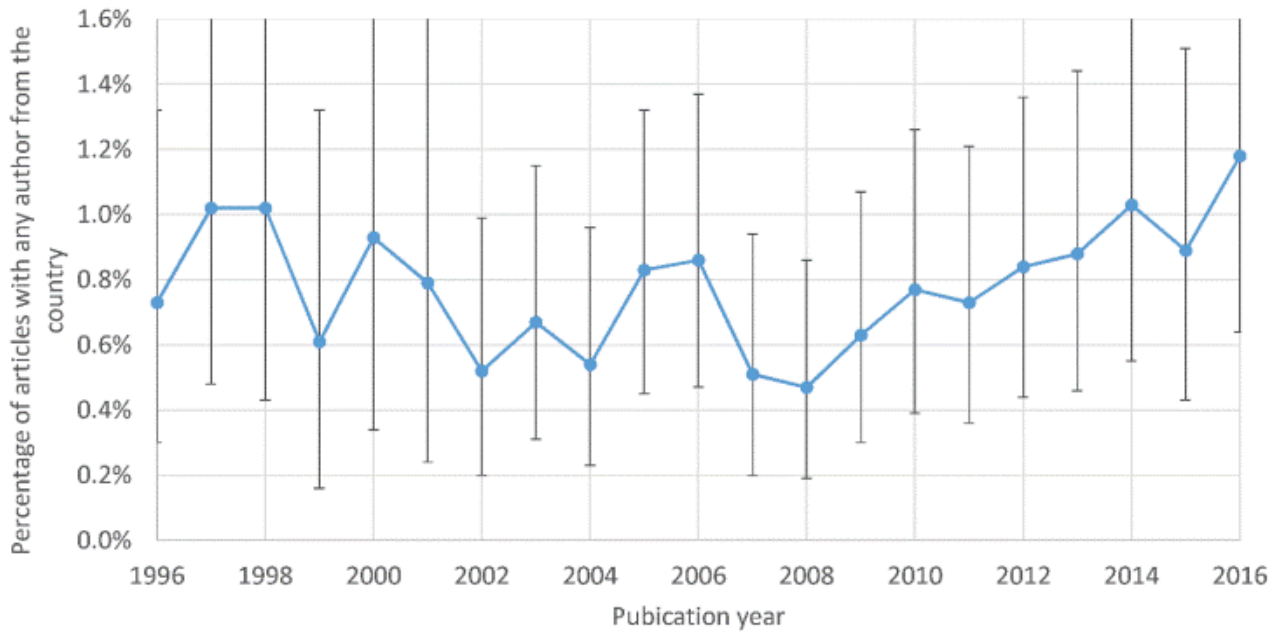

Figure 6. As Figure 4 for India.

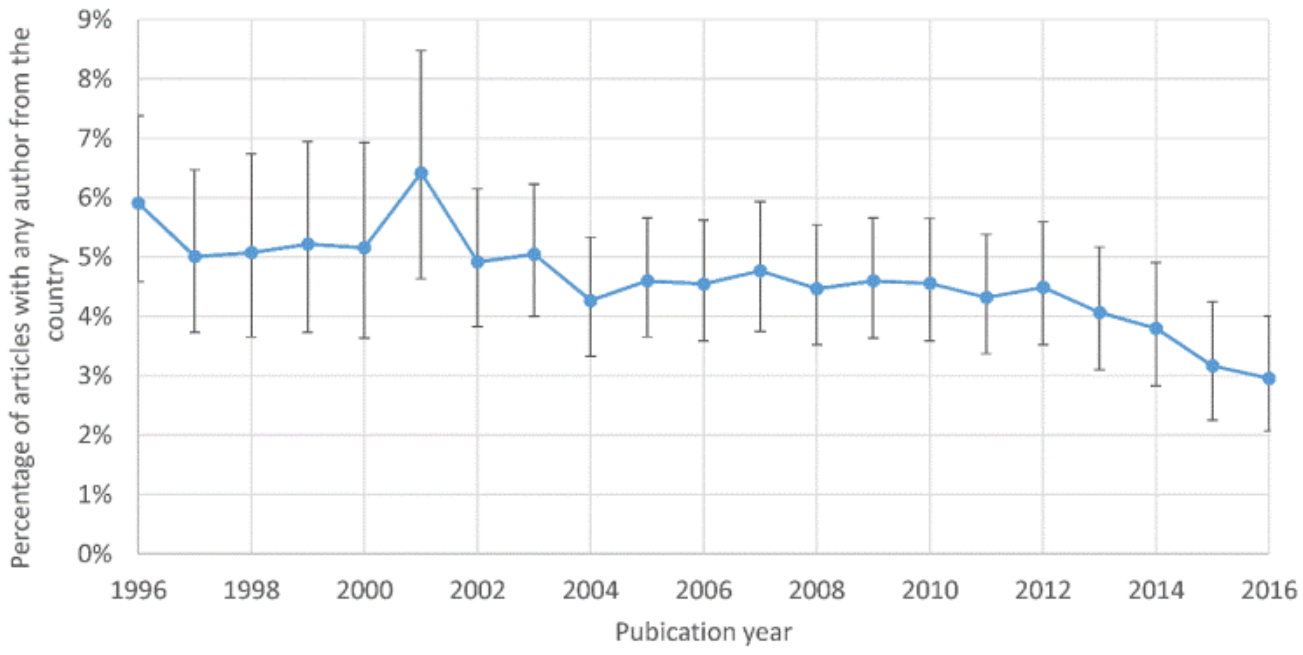

Figure 7. As Figure 4 for Canada.

## *4.2 Subject categories*

The percentages of times that the proportion of articles published by a country within a subject category falls within 95% prediction intervals from previous years (Figure 8, 9) are much lower than the corresponding percentages for individual journals (Figure 1, 2). There may be more substantial increasing or decreasing trends for country shares within subject categories than within individual journals caused by annual changes in the coverage of Scopus through adding, subtracting or re-classifying one or more nationally-focused journals. For example, Scopus had indexed 645,268 documents within journals containing “Chinese” in their title on 27 September 2017. For every article with a U.S. affiliation there were 41 affiliated with China. In contrast, Scopus had indexed 2,361,522 documents within journals containing “American” in their title on 27 September 2017. For every article with a China affiliation there were 42 from the U.S.A., confirming huge international biases in journals. Adding or removing such nationally biased journals to a category would affect statistics for all countries except any that contributed a similar share to their contribution to the remainder of the category.

China has the lowest line in Figure 8, which is partly due to increasing national research output and increasing coverage of Chinese journals. As Scopus has become more international, indexing increasingly many non-English periodicals (e.g., Spanish, Chinese), it has become less dominated by the English language journals that are the predominant publishing outlet of the UK, USA and Canada, decreasing their overall shares. Another contributory factor is that prediction intervals for categories will be narrower than prediction intervals for journals due to larger data sets. In consequence, small changes are more likely to result in values outside of the narrower prediction limits.

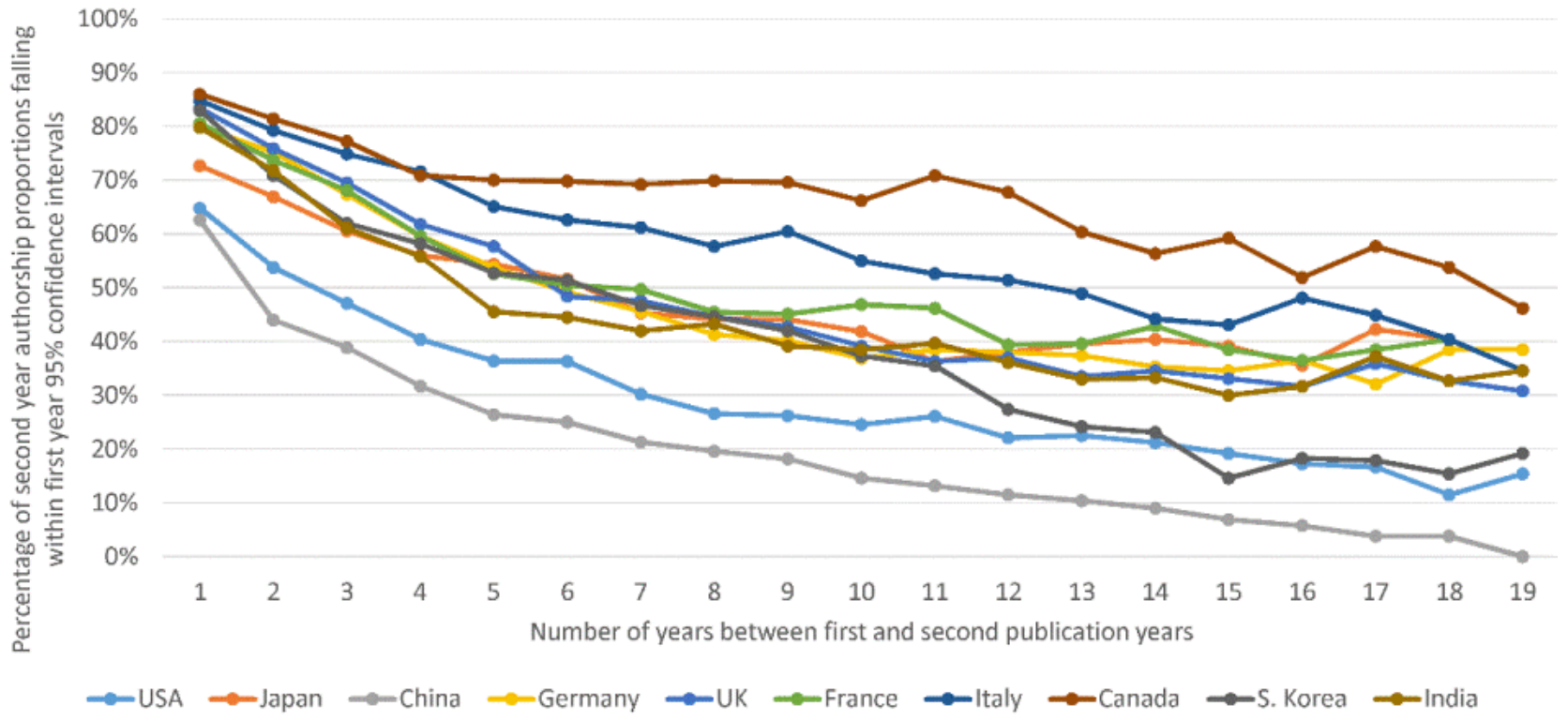

Figure 8. As Figure 1 except for 26 Scopus subject categories. There are 454-493 subject/year pairs for a 1 year gap falling linearly to 20-25 subject/year pairs for a 19-year gap. The raw data comprises n=47,067 individual tests.

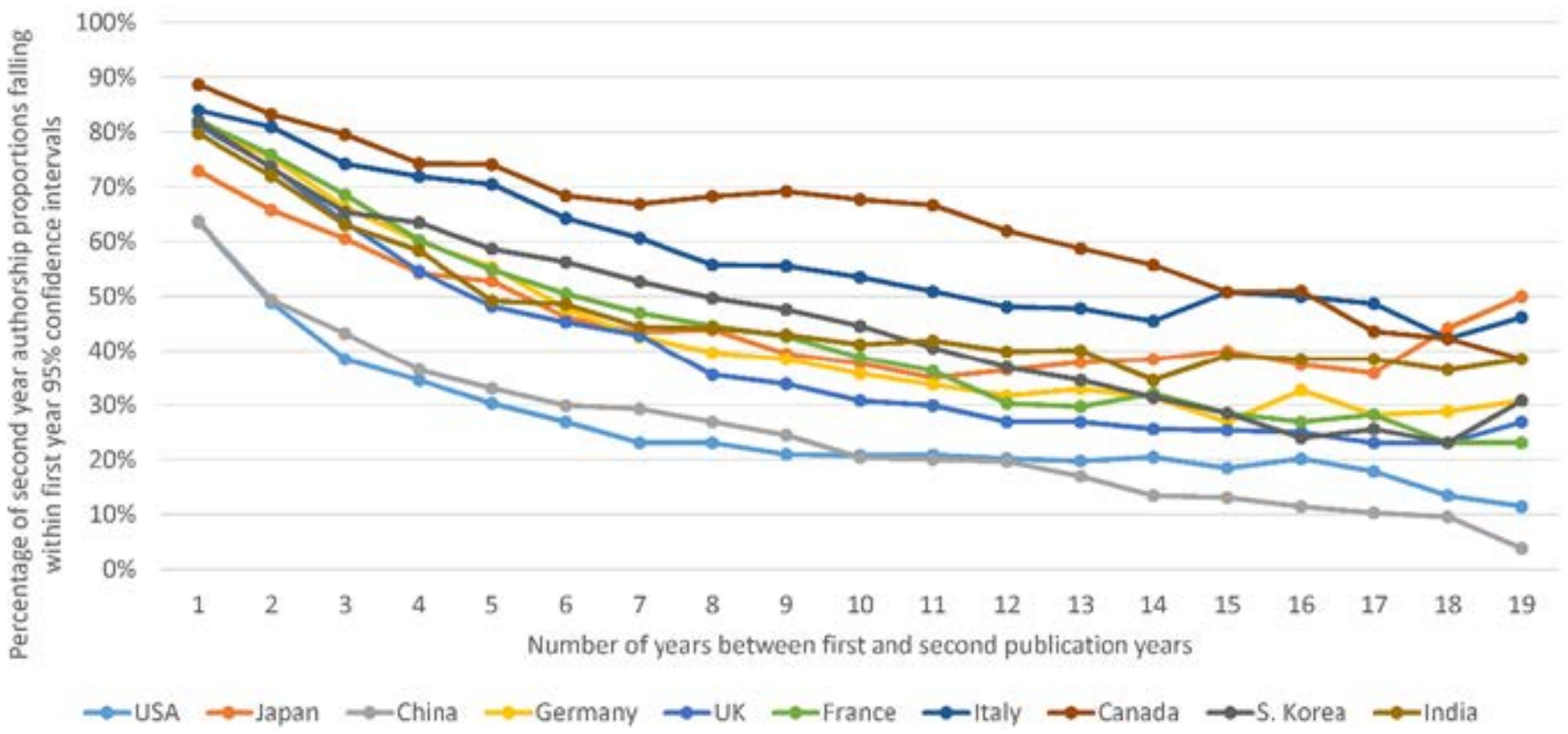

Figure 9. As Figure 8 except that the data for each country is for articles with *all* authors from that country.

Trends in the first category, Forestry, illustrate some key issues (Figures 10-13). The Scopus Forestry data is complete in all years, with the largest number of articles being 8016, from 2015 (i.e., below the threshold of 10,000 that causes data loss). For both Canada and the USA, there is a strong downward trend which accounts for the relatively low percentage of proportions from one year falling within 95% prediction intervals for the previous year.

There was a jump in 2006 for China, which is due to Scopus starting to index *Frontiers of Forestry in China* in this year. The decrease in 2011 was mainly due to the Scopus decision to cease indexing *Plant Physiology Communications*, which is published in China (http://www.oriprobe.com/journals/zwslxtx.html) and had 158 solely Chinese-authored articles in 2010.

For India, there was a sharp increase in share of Forestry in 2008, the year when Scopus started to index *Journal of Agrometeorology*. This international journal is published by the Association of Agrometeorologists, which formed in Gujerat, India (http://agrimetassociation.org/Default.aspx) and started publishing the journal in 1999 (http://agrimetassociation.org/Journal.aspx). When first indexed by Scopus in 2008, 128 out of the 151 articles had exclusively Indian authors. This more than doubled the 58 Indian-authored articles in the remaining Forestry journals combined. The sharp jump in Figure 12 is therefore entirely due to a Scopus indexing decision.

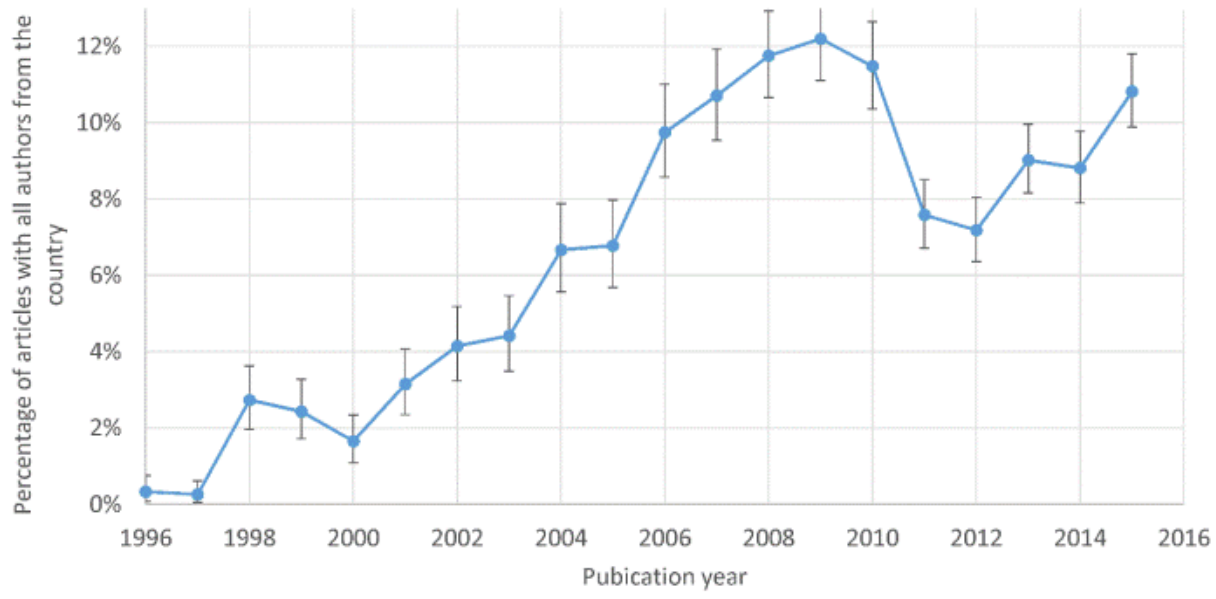

Figure 10. Prediction intervals for the probability (expressed as a percentage) that each article in the Forestry Scopus category have all authors from China. Each prediction interval assumes that the overall number of articles in the journal does not change over time.

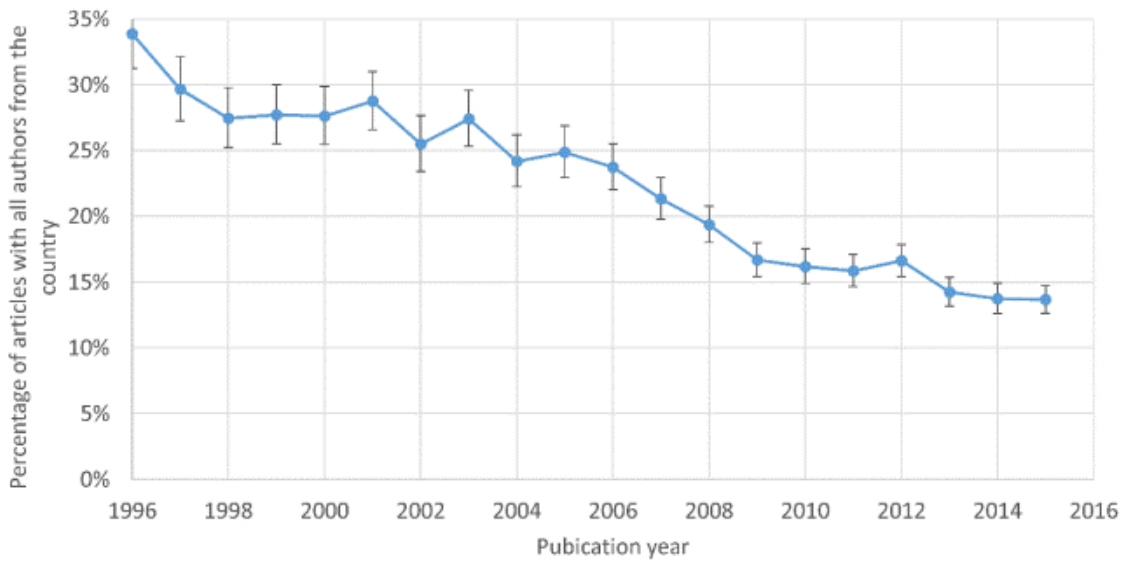

Figure 11. As Figure 10 for the USA.

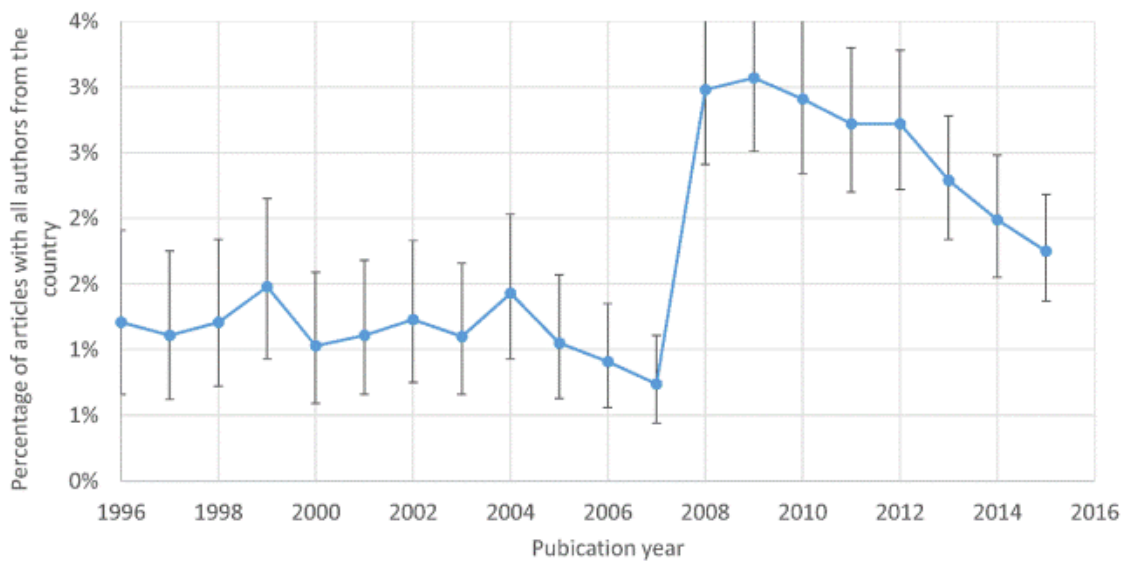

Figure 12. As Figure 10 for India.

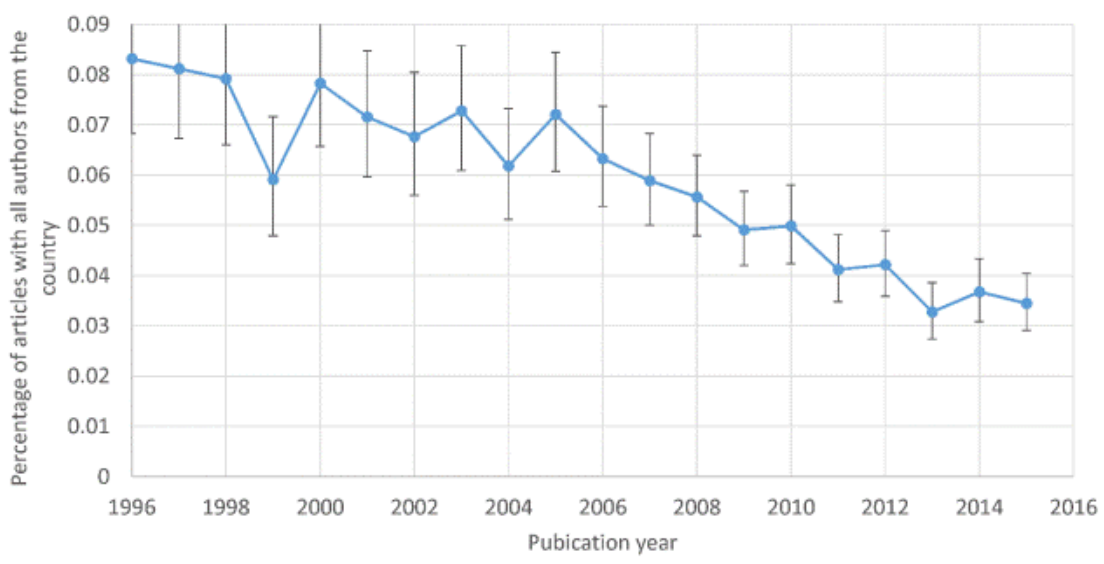

Figure 13. As Figure 10 for Canada.

## *4.3 Institutions*

The percentages of times that the proportion of articles published by an institution within a subject category falls within 95% prediction intervals from previous years (Figure 14, 15) are higher than the corresponding percentages for entire countries (Figure 8, 9) and for individual journals (Figure 1, 2).

The higher values (heights of lines in Figure 14, 15) compared countries for the same set of subject categories (Figure 8, 9) is counterintuitive because departments (or at least institution/subject combinations) seem to be inherently less stable than national output in a field (i.e., country/subject combinations). The increased stability is due to the proportions calculated being much smaller for institutions than for countries. Low numbers of publications in a category occur when an institution does not have a department specialising in the research area. With only a few articles expected each year, underlying trends are harder to detect because of the discrete nature of the data. For example, the expected whole *number* of publications to be published by an institution in each year may not change in response to a small change in the underlying production capability.

The percentages are implausibly close to 95% and above 95% for some countries and especially for short gaps in Figure 14 and 15. This is because the confidence intervals are necessarily conservative. For continuous data, confidence or prediction intervals can be exact but for discrete data they tend to be conservative. For example, if the prediction is the range 1-3 publications for a 95% prediction interval then this might give 99% confidence because any narrower range (e.g., 1-2 or 2-3) might give a confidence below 95%. Whilst this issue applies to all discrete data, it is largest for low numbers, as for the institution data here.

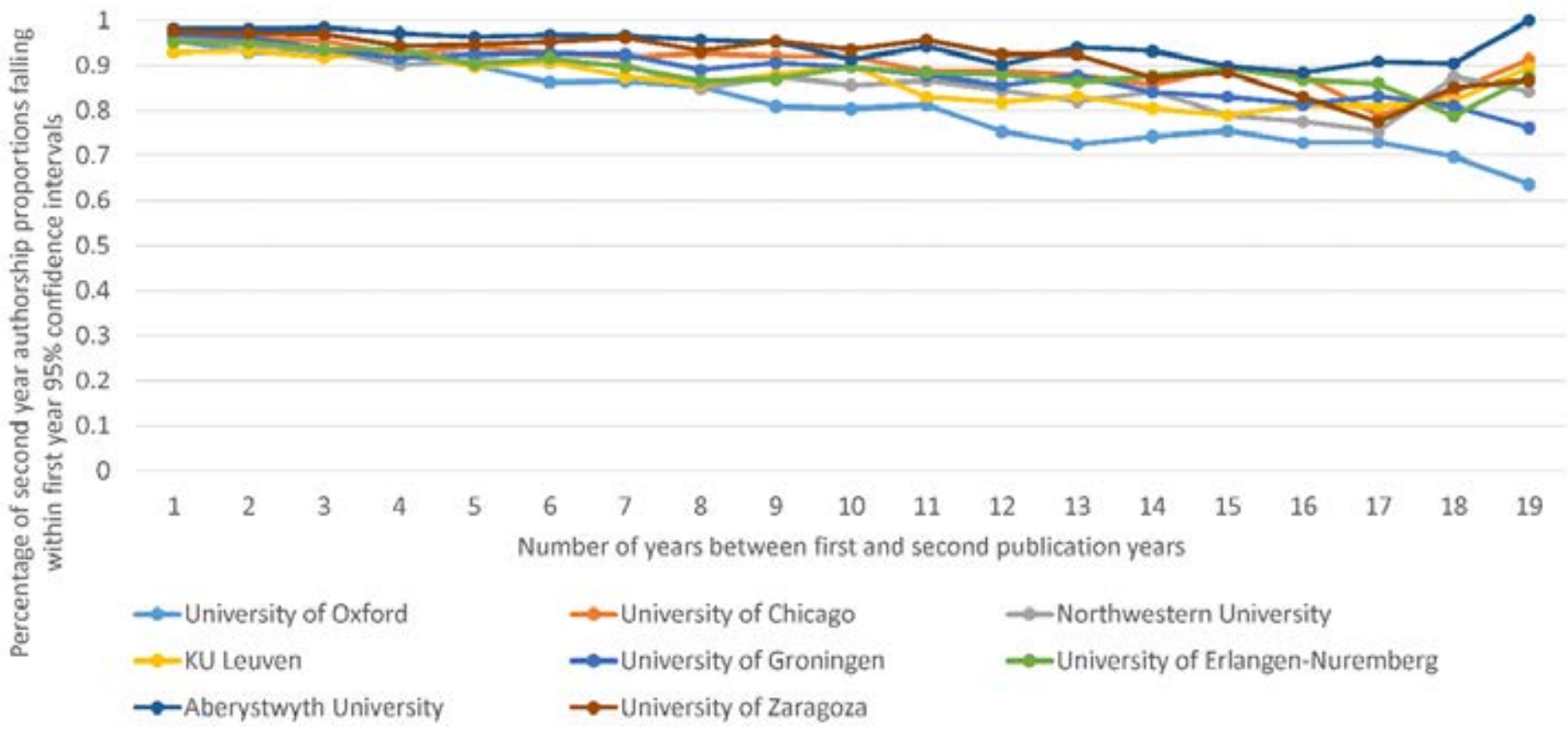

Figure 14. As Figure 8 except for seven varied ranking institutions.

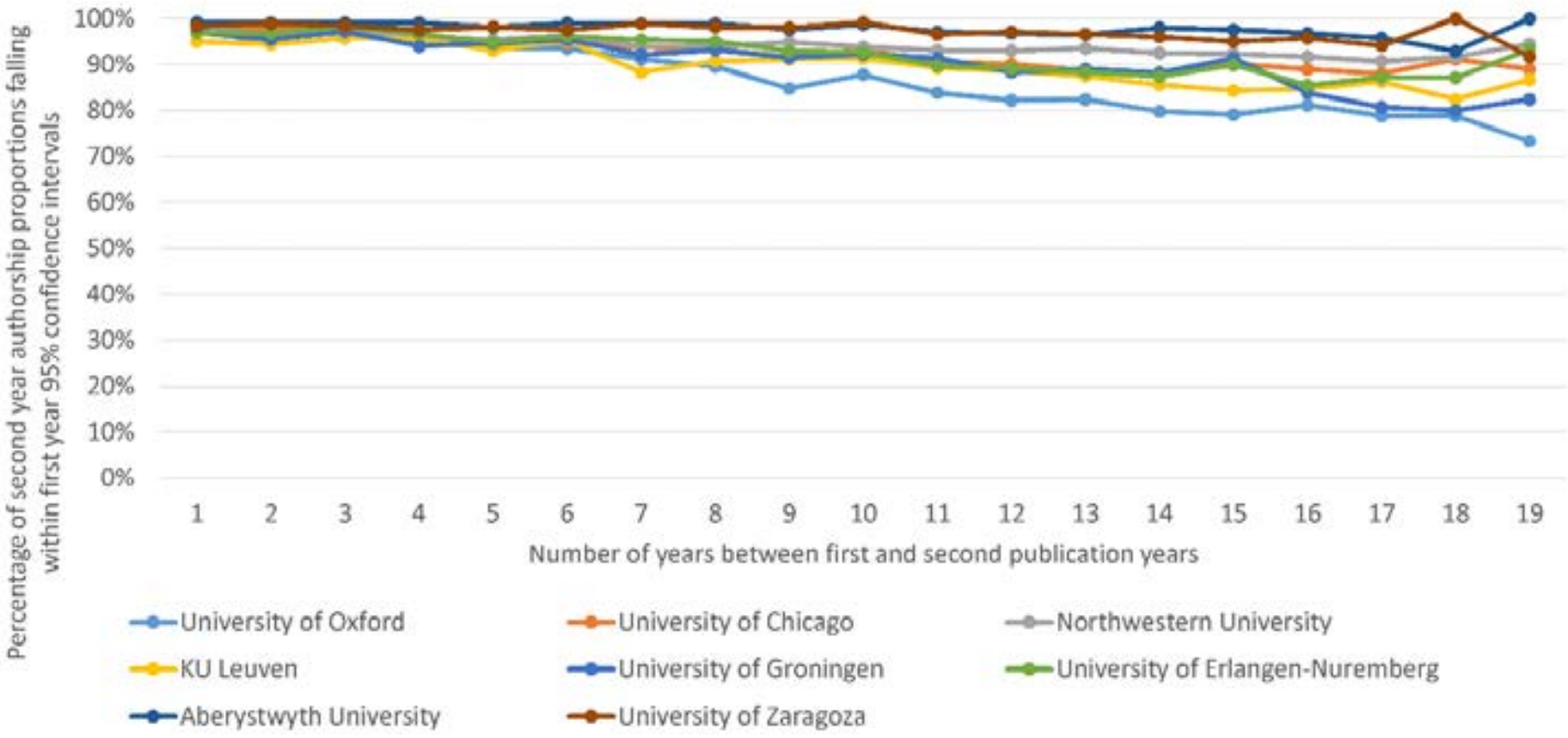

Figure 15. As Figure 14 except that the data for each university is for articles with *all* authors from that institution.

As an example of an institution/subject combination with relatively many publications, Pharmaceutical Science in KU Leuven (Figure 16) shows three spikes that fall outside of the prediction limits. The 1998 anomaly was due to 28 articles in Journal de Pharmacie de Belgique from 1998 in comparison to 3 the year before. This journal has been indexed in Scopus since 1949 with a maximum of 56 articles per year in all years except 1998, when it included 348. Some seemed to be research-based news instead of primary research (e.g., “Bioanalytical support in first-dosing-in-man studies”). The extra papers were from the Drug Analysis ’98 Symposium, Brussels, Belgium, May 1998, which the journal chose to publish. The 2008 and 2014 spikes were caused by increases in Bioorganic and Medicinal Chemistry that cannot be accounted for by Scopus’ coverage of the journal. In 2008 all except one of the articles had Erik De Clercq as an author or co-author. According to Scopus, 2008 was a below average year for production for De Clercq, so his (or his team’s) publications in Bioorganic and Medicinal Chemistry may represent a temporary shift in focus (in 2008 he received a prestigious lifetime achievement award for “Contributions to antiviral therapy”, http://www.epo.org/learning-events/european-inventor/finalists/2008.html). In 2014, most KU Leuven Bioorganic and Medicinal Chemistry articles were co-authored by Jan Balzarini, and 2014 was an above average production year for him (66 Scopus-indexed documents in comparison to 57 in 2013). The data set was also incomplete for 2014 (due to reaching the maximum of 10000 articles in the year and subject category) so both may be partial factors.

Representing an institution/subject combination with little data, Health, Toxicology and Mutagenesis in the University of Zaragoza (Figure 17) has one or two articles in most years and the prediction intervals suggest that the erratic pattern may be due to random variations rather than underlying trends.

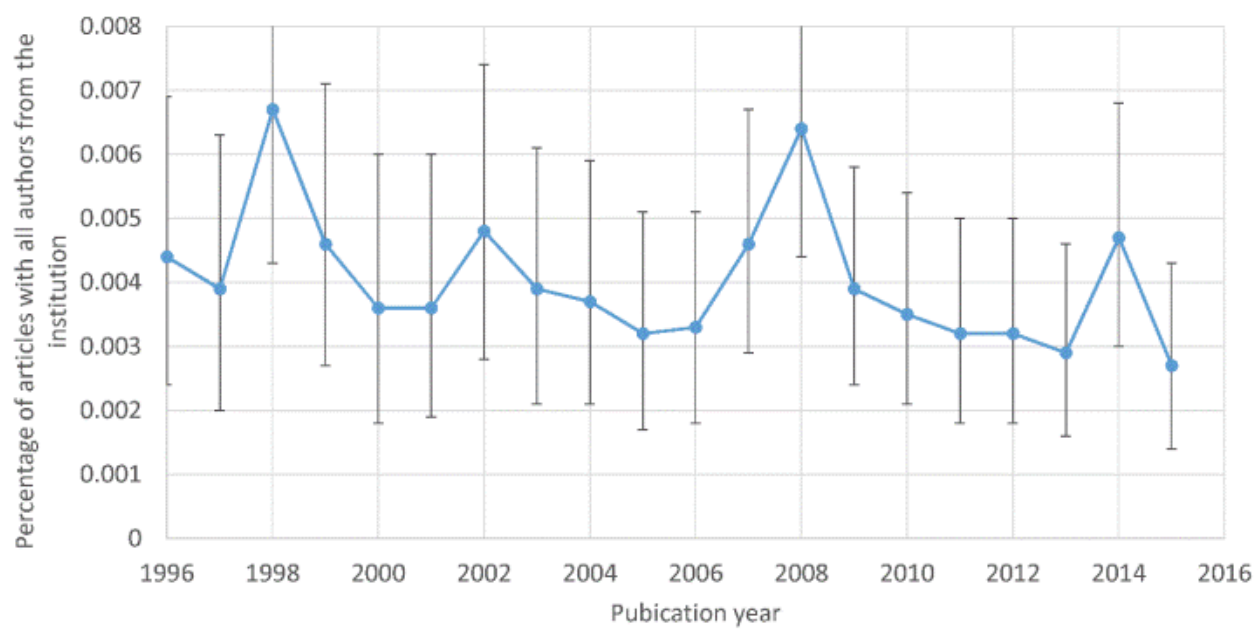

Figure 16. Prediction intervals for the probability (expressed as a percentage) that each article in the Pharmaceutical Science Scopus category have all authors from KU Leuven.

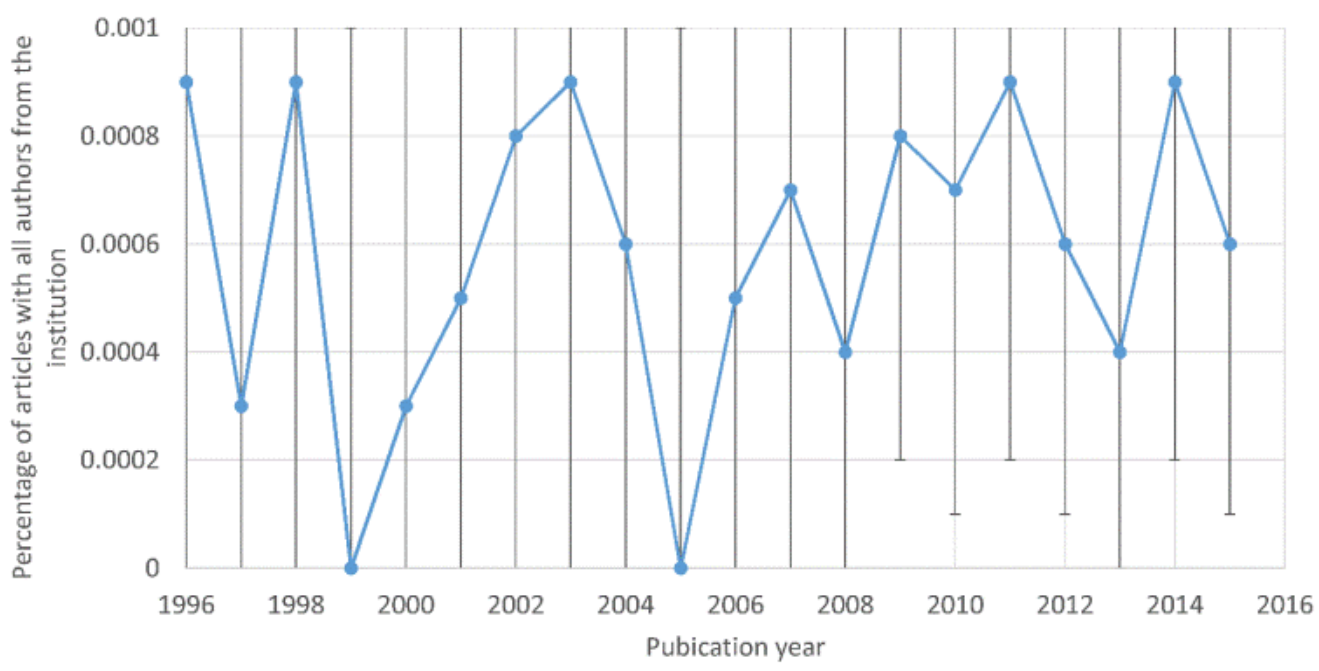

Figure 17. Prediction intervals for the probability (expressed as a percentage) that each article in the Health, Toxicology and Mutagenesis Scopus category have all authors from the University of Zaragoza.

# 5 Limitations

The results are limited through being tested primarily on large monodisciplinary journals, which are exclusively from the natural sciences and medicine, with a bias towards medicine and physics. The conclusions should therefore be interpreted as speculative when applied to the social sciences, arts and humanities. The primary results are also restricted by being applied to journals rather than subject categories, which are the more natural unit of assessment for practical research evaluations and theoretical investigations of science (e.g., Falagas, Karavasiou, & Bliziotis, 2006; Vergidis, Karavasiou, Paraschakis, Bliziotis, & Falagas, 2005). The decision to use only journals for the primary analysis was due to the variability in coverage of subject categories, which would give less clear results. Thus, the evidence here is for the plausibility of the prediction intervals based upon the theoretical discussion and the empirical evidence but, as in almost all kinds of mathematical modelling, it should not be taken as proving that the model is "correct".

The results for subject categories are limited by the inclusion of a restricted range and the methodological decision to include the entire Scopus subject category in each case, allowing non-systematic sources of variation from changes in the journals indexed by Scopus. In practice, an assessment of changes over time would need to ensure that the same journals are used throughout the period and that the subject area being analysed is covered comprehensively. Taking this extra variability into account, the country-level subject results are broadly consistent with the research hypotheses.

The results for seven institutions may not apply to much smaller institutions or those with much lower research production. These are also limited by the set of subject categories chosen.

# 6 Conclusions

The arguments and empirical evidence suggest that it is very broadly reasonable to use prediction intervals to estimate the likely number of future outputs for a set of researchers, whether a country, institution or department, from the number of articles published in a journal or subject area, although only countries and large monodisciplinary journals were included in the primary analysis. It is also broadly reasonable to use confidence intervals to estimate the underlying production capability of a set of researchers. This is because the prediction intervals tested above are based on the same assumptions as confidence intervals, but generate wider intervals to account for the random variance in the publications from the second year. Prediction or confidence intervals could be used by research managers and policy makers monitoring the progress of a group of researchers and wishing to know whether underlying research capacity has expanded, contracted or stayed the same relative to world capacity, irrespective of the number of articles published each year. This should be used in conjunction with other output indicators (e.g., absolute output, publication efficiency) as well as impact or quality based assessments to give a multidimensional understanding of changes over time. Managers may decide that it is more appropriate to compare against a specific set of other researchers (e.g., a set of economically comparable countries rather than the rest of the world) and for this could discard other publications before starting the analysis.

An important caveat to the above is that strong conclusions should not be drawn about statistical evidence from the production capability model. This is because its assumptions are a simplification of the situation in practice and in any case all aspects of the world science system are continually changing. Thus, statistical evidence should be perhaps interpreted as *permission* to consider the possibility that there has been a change in the group analysed, or a difference between the two groups compared, but not proof of this. It cannot be proof because of the influence of system-level changes that the model does not consider, such as disproportionate expansion in the coverage of a country's journals by the citation index used. In this context, analysts may wish to consider whether the extra information given by prediction intervals outweighs the confusion that may be created by the additional complexity of the information presented for the managers that will read their report.

If considering a range of indicators (e.g., publication proportions for a set of broad fields) then prediction limits could be used from the opposite perspective from above: to give policy makers *permission* to ignore differences that are not statistically significant. Their task would then be to analyse the statistically significant differences qualitatively to assess whether the differences were (a) large enough to be important and (b) likely to be caused by factors under their control rather than database coverage or competitor changes. This system is not fool proof because there might be substantial changes that are not flagged as statistically significant because they are masked by database coverage changes but it would at least be a simple and transparent method to narrow down the number of likely important differences.

Confidence or prediction intervals will be unhelpful in situations where research managers know that the system has changed during the time interval examined. For example, if a country closes or opens several new universities or a research group expands or contracts substantially then the confidence or prediction intervals will tell the manager nothing because the system assessed is already known to have changed. Thus, confidence or prediction intervals are most useful in situations where the size of a system is reasonably static. For example, in a static system, confidence or prediction intervals may help governments to decide whether policy changes (e.g., reorganising the funding system) have influenced research output relative to the world norms, if other system changes are minor.

Although the experiments have focused on temporal changes, confidence or prediction intervals can also be used to compare different groups of researchers at the same point in time, if both are covered equally by the academic database used. Such comparisons are much more difficult than quality or impact comparisons because research groups, institutions and countries rarely have the same sizes as each other. Moreover, it is often difficult to size normalise groups (e.g., dividing publications by faculty: Toutkoushian, Porter, Danielson, & Hollis, 2003) because this requires accurate and comparable data about personnel and perhaps also funding for a fair comparison.

As a concrete example about how confidence or prediction intervals may be used in practice, suppose that a government changes its method of allocating research funding but not the amount of funding and wishes to know whether research output increased as a result (it could test this before testing for changes in citation impact). Since publishing volume increases annually, the total number of papers produced is likely to have increased, irrespective of whether the policy had been successful. A reasonable strategy would be to plot a graph of share of the world's research over time, together with error bars denoting confidence or prediction limits, as calculated above. This graph would help to show whether there was statistical evidence of an underlying change in world share after the policy change. Alternatively, the government might consider only comparable nations and plot instead a graph of share of publications relative to this set. This would remove global influences, such as the rise of China, that might be irrelevant.

Although the confidence or prediction intervals and theoretical model are designed for shares of the world's output, they can also be applied to absolute numbers of publications by multiplying each year's values (proportion and both confidence or prediction limits) by the world total number of publications in that year. The underlying logic of the research production capability model would still apply but the figures would also reflect the expansion in publishing (and indexing in citation databases). This would have the advantage of being more intuitive but may tend to give a more optimistic message than the data warrants, given that publishing expansion is normal.

Finally, if confidence intervals are calculated using Wilson's score interval then they should be interpreted with the understanding that they are for the underlying probability to publish in a journal or subject rather than being a prediction about the likely range of future proportions, if there has been no underlying change. Thus, 95% confidence intervals should not therefore be interpreted as predictions for the following year's empirical value, in the absence of change. It would be reasonable to interpret them as 95% confidence intervals the future year's *underlying probability in the absence of change* but not of the future year's *empirical value*. In the absence of change, future year empirical values could be expected to fall within 95% confidence intervals less than 95% of the time, with the exact percentage depending on the sample size and variability: prediction intervals are better for this case. For applications to subject categories, care should be taken to ensure that the set of journals is consistent between the years assessed to avoid spurious findings.

## Appendix A: Journal list

Applied Mathematics & Computation; Applied Physics Letters; Applied Surface Science; Astronomy & Astrophysics; Astrophysical Journal; Biochemical & Biophysical Res. Comm.; Biochemistry; Bioorganic & Medicinal Chemistry Lett.; Brain Research; Cancer Research; Chemical Physics Letters; Geophysical Research Letters; Inorganic Chemistry; J. of Agricultural & Food Chemistry; J. of Applied Physics; J. of Applied Polymer Science; J. of Biological Chemistry; J. of Chemical Physics; J. of Immunology; J. of Neuroscience; J. of Organic Chemistry; J. of Power Sources; J. of the American Chemical Society; J. of Virology; Japanese J. of Applied Physics Part 1; Langmuir; Macromolecules; Materials Science & Eng. A; Monthly Not. R. Astronomical Soc.; Nuclear Instruments & Meth. Physics A; Physica B Condensed Matter; Physical Review A.; Physical Review Letters; Tetrahedron; Tetrahedron Letters; Thin Solid Films.

## Appendix B: Category list

Forestry; History and Philosophy of Science; Cell Biology; Organizational Behavior and Human Resource Management; Fluid Flow and Transfer Processes; Spectroscopy; Computer Vision and Pattern Recognition; Statistics Probability and Uncertainty; Geology; Finance; Fuel Technology; Control and Systems Engineering; Health, Toxicology and Mutagenesis; Applied Microbiology and Biotechnology; Polymers and Plastics; Discrete Mathematics and Combinatorics; Dermatology; Transplantation; Endocrine and Autonomic Systems; Emergency Nursing; Pharmaceutical Science; Atomic and Molecular Physics and Optics; Social Psychology; Human Factors and Ergonomics; Small Animals; Medical Laboratory Technology

# Appendix C: Analysis steps

1. Get raw publication data: tab-delimited plain text file of publications with one column containing country affiliations of authors (e.g., from Scopus). One file per journal or subject category, containing all years.
2. Split the data by year so that there is a separate file for each publication year. File names must have no spaces then end in [space][year]-world.txt. E.g., **0003-6951 2000-world.txt**. With Webometric Analyst, this requires *Tab-sep|Split file into multiple parts based on Column n*. Select enter the number of the column containing the year and, if the data is in year-month-day format, also specify hyphen – as the text to ignore from. To rename all files by adding -world at the end, use Webometric Analyst *File|Rename all files in folder (including path) by replacing one string with another*.
3. Create a second file with at least one author from the given country, naming these files as above but with -Any[countryname].txt at the end. E.g., **0002-7863 2011-world-AnyCanada.txt**. In Webometric Analyst, use *Citations|Copy Scopus records when one or all authors are from a given country*. Enter the name of the country in the same format used in the text file. Specify at least one or all authors.
4. Make reports (not needed) and AllData files with *Reports|Calculate MNLCS, gMNCS and EMNPC (NPC) for a set of any tab-delimited files (one or more files, each processed separately).*
5. Rename the AllData files to [Any|Only][countryname] AllData.txt. e.g., AnyIndia AllData.txt or OnlyUK AllData.txt.
6. Move the renamed AllData files to a single folder.
7. Create the confidence intervals using *Citation|Test confidence interval in publication time series in multiple AllData files with year at end of filename*.